\documentclass{mem}
\usepackage{natbib}\usepackage{txfonts}\usepackage{balance}
\usepackage{graphicx}
\usepackage[a4paper,breaklinks,dvipdfm]{hyperref}
\idline{84}{1}
\usepackage{txfonts} 

\begin{document}
\def\teff{$T\rm_{eff }$}
\def\kms{$\mathrm {km s}^{-1}$}

\def\la{\mathrel{\hbox{\rlap{\hbox{\lower4pt\hbox{$\sim$}}}{\raise2pt\hbox{$<$}}}}}
\def\ga{\mathrel{\hbox{\rlap{\hbox{\lower4pt\hbox{$\sim$}}}{\raise2pt\hbox{$>$}}}}}

\title{
The sub-classes of ultraluminous X-ray sources
}

   \subtitle{}

\author{
Jeanette C. Gladstone\inst{1} 
          }

  \offprints{J. C. Gladstone}

\institute{
Avadh Bhatia Fellow, University of Alberta, Department of Physics, Edmonton, Alberta, T6G 2E1
\email{j.c.gladstone@ualberta.ca}
}

\authorrunning{Gladstone}

\titlerunning{sub-classes of ULXs}

\abstract{

Ultraluminous X-ray sources (ULXs) have been an enigma since their discovery. \emph{ASCA} showed that they are accreting black holes with unknown mass, while \emph{Chandra} \& \emph{XMM} data shows that  many of the lower luminosity ULXs can be explained by extreme accretion onto stellar remnant black holes. The discovery of hyperluminous X-ray sources has led to speculation that we may, once again, have found the elusive intermediate mass black holes, yet in the last year, a third sub-class of \emph{extreme} ULXs has emerged, with their nature, as yet, unknown. Here we define and discuss the sub-classes of these fascinating systems, exploring possible explanations for their nature, and indicating possible future steps in their analysis. 

}
\maketitle{}

\section{Introduction}

For more than 30 years, bright	($L_X \ga 10^{39}$ erg~s$^{-1}$), extra-galactic, non-nuclear X-ray point sources have been studied in an effort to explore their nature. They are known to be black holes, but their luminosities preclude the possibility of stellar mass black holes (sMBHs; $3\la M_{BH} \la 20 M_\odot$) accreting isotropically below the Eddington limit, while their positions within their host galaxies - along with dynamical friction arguments Ð also rule out super-massive black holes (SMBHs; $M_{BH} \ga 10^6 M_\odot$) as a possible source of emission.

The simplest explanation is that they contain intermediate mass black holes (IMBHs; $\sim 10^2 \la M_BH \la 10^5 M_\odot$; e.g., Colbert \& Mushotzky 1999) accreting in a known, sub-Eddington, accretion state. IMBHs would provide a missing link in the mass-scaling of black holes, and are of great cosmological importance, e.g. as the building blocks of SMBHs (Volonteri 2010), whose mass would subsequently build up through accretion of gas (e.g. Yu \& Tremaine 2002), and/or hierarchical black hole mergers (e.g. Schneider et al. 2002). As a result, they would also have significant implications on our understanding of the formation/evolution of their host galaxies (e.g. Madau \& Rees 2001; Ebisuzaki et al. 2001). 

\begin{table*}
\caption{The sub-classes of ultraluminous X-ray sources}
\label{tab:sub-class}
\begin{center}
\begin{tabular}{lll}
\hline
Sub-class					& Abbreviation		& Luminosity range 	\\
						&				& (erg s${-1}$)		\\
\hline
standard or low luminosity 			& sULXs	& $\sim 10^{39}$ to $\sim 2 \times 10^{40}$	\\
ultraluminous X-ray sources			&		&	\\
extreme ultraluminous X-ray sources	& eULXs	& $\sim 2 \times 10^{40}$ to $\sim 10^{41}$	 \\
hyperluminous X-ray sources			& HLXs	& $ \ga 10^{41}$	\\
\hline
\end{tabular}
\end{center}
\end{table*}

The alternative is that we are observing massive stellar-remnant black holes (MsBHs;  $20 \la M_{BH} \la 100 M_\odot$;  Feng \& Soria 2011) that are either breaking (e.g. Begelman 2002) or circumventing (e.g. via geometric beaming, King et al. 2001, and/or relativistic beaming, K{\"o}rding et al. 2002) their Eddington limit. Such massive objects are still formed as the end product of a single, current-generation star, however, the metallicity of their environment must be much lower than solar (e.g. Fryer \& Kalogera 2001). If these sources are breaking the Eddington limit, they would be the most extreme accretors in the local universe, requiring new accretion physics which provides possible clues for understanding the rapid build-up of the first quasars.

In this paper, we discuss the nature of these objects, and how observations of ULXs have challenged our understanding of black holes and accretion physics. We show how recent findings have led to subdivision of ULXs, and explain the current consensus about their nature. 

\section{The sub-classes of ULXs}
\label{section: sub-class}

In an effort to understand these sources, much work has gone into the study of ULXs. Many ULXs emit in the $\sim$$10^{39}$ to $\sim$2$\times$$10^{40}$ erg s$^{-1}$ range, with research showing that many appear to be sMBHs and MsBHs accreting at or above the Eddington limit. In 2004, a new sub-class was formed, termed the hyperluminous X-ray sources (HLXs; e.g. Matsumoto et al. 2004), that are thought to be amongst the strongest IMBH candidates. In the last year a third group has emerged, the new \emph{extreme} ULXs, with the nature of these sources still appearing very uncertain. Each of these sub-classes are defined in Table \ref{tab:sub-class}.


\subsection{Standard or low-luminosity ULXs}

These sources emit in the range of $\sim10^{39}$ to $\sim2\times10^{40}$ erg s$^{-1}$, and have been heavily studied in multiple wavebands.

\subsubsection{X-ray analysis of standard ULXs}

Early X-ray spectral studies showed that sULXs were well fit by a cool disc ($\sim$ 0.2 keV) plus power-law, where the fit implied the presence of a $\sim 2 \times 10^4 M_\odot$ black hole (an IMBH; Miller et al. 2003). However, subsequent re-analysis detected a break in the power-law continuum at  $\sim$ few keV - a feature not seen in the standard sub-Eddington states (Stobbart et al. 2006; Gladstone et al. 2009). As a result, this combination of both a soft excess (modelled by a cool disc) and a spectral break above $\sim$ 3 kev have become the criteria for the new \emph{ultraluminous} state (Gladstone et al. 2009).

Temporal studies of sULXs show that large scale (up to $\sim$ order of magnitude) variability is present on timescales of days/weeks (e.g. NGC 5204 X-1, Roberts et al. 2006) to years (e.g. ULX population of NGC 4485 \& 4490, Gladstone \& Roberts 2009). Intra-observational analysis reveals that sULXs have suppressed short-term variability, with many showing levels an order of magnitude lower than typical for sMBHs and AGNs (Heil et al. 2009). This is consistent with the picture put forward by Middleton et al (2011a), in which the ULXs are dominated by disc emission (albeit not fully-thermalised in many cases), or a large photosphere in the 0.5-10.0 keV bandpass. 

X-ray analyses seem to be pointing towards a population of sMBHs \& MsBHs accreting at or above their Eddington limit. However, two objects (M82 X42.3+59 \& NGC 5408 X-1) show quasi-periodic oscillations (QPOs). M82 X42.3+59 has been observed over $\sim$7 yrs, showing only a $\sim$factor 2 variation in flux, with a 3-4 mHz QPO during its brightest observation (Feng et al. 2010). By applying a mass scaling, the authors found  $M_{BH}\sim$12,000-43,000 $M_\odot$. NGC 5408 X-1 also shows little long-term variability, with intra-observational studies revealing a 10-40 mHz QPO (e.g. Strohmayer et al. 2007). Dheeraj \& Strohmayer (2012) combined this with spectral and timing analysis to find  $M_{BH} \sim$800 $M_\odot$ (see Middleton et al. 2011 for alternative interpretation). Each of these sources emit at $\sim10^{40}$ erg s$^{-1}$, and so are close to the eULXs class, and 
 could belong to a different population of sources harbouring IMBHs. The alternative is that we are observing MsBHs at a different inclination, with the presence of a wind and/or different accretion state. These sources require further study in order to (a) confirm the nature of the QPO, and (b) determine the physical origin of their features in order to explain their difference from the rest of the sULX population. 

By considering the near-by sULX population s a whole, Grimm et al. (2003) constructed an X-ray luminosity function (XLF) for the local universe, finding an unbroken luminosity function connecting sULXs to the standard X-ray binary population. Recent work by Swartz et al. (2011) confirmed this and the presence of a break in the XLF of high mass X-ray binaries (HMXBs) at $\sim 2 \times 10^{40}$ erg s$^{-1}$, providing the upper limit for this new sub-class.

\subsubsection{Optical follow-up of standard ULXs}

 These sULXs are found preferentially in actively star-forming galaxies (Kaaret 2005), starburst, interacting/merging and dwarf galaxies (Swartz et al. 2008), with low metallicities (Pakull \& Mirioni 2002; Zampieri \& Roberts 2009); environments suitable for HMXBs and the formation of MsBHs.

Potential stellar companions of sULXs tend to be blue ($m_{\rm V} \sim$ 20 -- 26; Roberts et al. 2008; Gladstone et al. 2013), suggesting O or B type stars, yet we must also consider possible contamination from the irradiated  accretion disc/stellar surface (e.g. Copperwheat et al. 2005; Madhusudhan et al. 2008). By accounting for the accretion disc, Gladstone et al. (2013) were able to impose inclination-dependent upper and/or lower limits on the black hole's mass for 7 sULXs. O stars could be ruled out in 10 of the 18 cases tested, while B stars were possible in all scenarios.

Spectroscopic follow-up has provided a blue, almost featureless spectrum (e.g. NGC 1313 X-2 \& Ho IX X-1, Roberts et al. 2011; NGC 5408 X-1, Kaaret \& Corbel 2009). This suggests the light is non-stellar in origin, and may be dominated by the accretion disc. Initial results from multi-epoch spectra of two sULXs (NGC 1313 X-2 \& Ho IX X-1) detected radial velocity variations; however, they may not be sinusoidal (Roberts et al. 2011). Further analysis, using stacked spectra shows that the He\textsc{ii} emission, used for this study, are slightly extended, implying that a nebular component is present (e.g. Ho IX X-1, Moon et al 2011; NGC 5408 X-1, Gris{\'e} et al. 2012). With current data, we are unable to statistically separate these features. Therefore, we must search for alternative stellar features in the optical and near infrared (NIR) to settle the debate on the masses of these black holes (Gladstone et al. \emph{in prep}). 

\subsection{Hyperluminous X-ray sources}
\label{subsection: HLXs}

Combining MsBHs with maximal super-Eddington radiation can explain ULXs with $L_X$ up to $\sim10^{41}$ erg/s, but not beyond. However, some exhibit luminosities above this - HLXs. They require exotic scenarios; e.g. stripped nuclei of dwarf galaxies (Bellovary et al. 2010), recently recoiled SMBHs (Jonker et al. 2010), tidal disruption events by SMBHs (Jonker et al. 2012). Similarly, IMBH X-ray binaries remain an intriguing possibility. With only two HLXs confirmed to date, and few other candidates having been identified, this is an avenue open to many future studies. 

\subsubsection{M82 X-1}

The first confirmed HLX is also the nearest, and was detected by some of the earliest X-ray satellite missions (\emph{Uhuru}, Forman et al. 1978; \emph{HEAO 1}, Piccinotti et al. 1982). M82 X-1 resides in a galaxy highly active in X-rays,  possibly due to a burst of star formation $\sim$$10^7$-$10^9$ yrs ago (Collura et al. 1994). Portegies Zwart et al. (2004) suggested that it was located in a young stellar cluster, however reanalysis showed that it is actually offset by 0.65 arcseconds from the cluster (Voss et al. 2011). 

X-ray spectral and timing analysis suggested $M\sim1000 M_\odot$ (Fiorito \& Titarchuk 2004), with a spectral evolution that is different from other ULXs (Matsumoto et al 2004). An apparent transition from the low hard state to the high soft state was observed by Feng \& Kaaret (2010). Combined with limits from radio \& infrared observations, this indicates a black hole mass range of $9\times10^4-5\times 10^5$ (Yuan et al. 2007; however alternatives have been suggested, e.g. Okajima et al. 2006). This apparent divergence from lower luminosity ULXs is strengthened by the presence of short term variability (Tsuru et al. 2004), with intra-observational analysis revealing a 50-100 mHz QPO (Strohmayer \& Mushotzky 2003), along with a break in the PSD at $\sim$ 34 mHz (Dewangan et al. 2006). Variability was also observed on longer timescales, with a 62 day periodicity 
suggested to be the orbital period 
 of the binary (Kaaret et al. 2006; Kaaret \& Feng 2007). 

\subsubsection{ESO 243-49 HLX-1}

This object is the brightest of the HLXs, peaking at $\sim10^{42}$ erg s$^{-1}$. ESO 243-49 HLX-1 displays large scale (approximately order of magnitude) long-term variability, 
with a periodic ($\sim$370 d; Servillat et al. 2011) outbursting cycle thought to be triggered by a companion in a highly eccentric orbit (Lasota et al. 2011).  During these outbursts the spectrum appears to change from a more power-law like slope to a more thermal shape, with the authors proposing a sub-Eddington state transition (Servillat et al. 2011). Hardness-intensity diagrams, showing the spectral evolution of this source, also echo those seen in Galactic systems at sub-Eddington rates (Farrell et al. 2011; Servillat et al. 2011). ESO 243-49 HLX-1 has a unique optical counterpart (Soria et al. 2010), observed at different stages of its 2010 outburst. \emph{HST} data was taken $\sim$1 month after the peak (Farrell et al. 2012), while VLT data was taken $\sim$2 months later (Soria et al. 2012). Farrell et al. (2012) found  a  blue spectral distribution, with no evidence of a Balmer break, with concurrent optical \& X-ray fitting giving a preference for a younger stellar population. Soria et al. (2012) reported a drop of  $\sim$1 mag in each optical band, while the X-ray flux dropped by a factor $\sim$2. This drop appeared to confirm the presence of multiple optical emission components; a (non-variable) population of stars, the outer region of an accretion disc, and/or an exceptionally luminous donor star. This suggests that any stellar cluster light is not dominant at outburst maximum, while fitting suggests the requirement for an irradiated disc during these bright periods. However, additional studies are required to both confirm the drop and decipher the optical components. 

\subsubsection{Others in this class}

Sutton et al. (2012) identified three additional candidates that also displayed spectra and long-term variability more similar to sub-Eddington states, suggesting the presence of IMBHs. One has also been observed in the Cartwheel galaxy (N10, Pizzolato et al. 2010) and another is labelled CXO J122518.6+144545 (Jonker et al. 2010). Much additional work needs to be done to both confirm that these sources are HLXs, and to explore their nature, while more of these objects should be identified if we are to statistically exploit this new sub-class of objects. 

\subsection{Extreme ultralumious X-ray sources}

This third sub-class of ULXs has only emerged in recent months. These are objects with X-ray luminosities of $\sim$2$\times$$10^{40}-\sim$$10^{41}$ erg s$^{-1}$. The range arose from a combination of the break in the XLF in our local universe and the limits on extreme MsBH accretors, leaving the few sources between these two limits to be classified as eULXs. 

Sutton et al. (2012) found that the X-ray spectral and temporal analysis of eULXs revealed a marked divergence from sULXs. eULXs are typically harder, with an absorbed $\Gamma \sim$ 1.7 (in contrast, ULXs in Gladstone et al. (2009) and Berghea et al. (2008) show $\Gamma \ga$ 2 below $10^{40}$ erg s$^{-1}$). While more investigations are required to explore this new class of objects, two of the sample are discussed below. 

\subsubsection{NGC 1042 ULX}

NGC 1042 ULX was first discussed in Sutton et al. (2012), where the X-ray spectral analysis revealed a power-law like spectrum with $\Gamma \sim$1.5, and temporal studies revealed the presence of short-term variability. This combination is more comparable to the low hard state, than the ultraluminous state. As a result this may be another candidate IMBH. 

\subsubsection{NGC 5907 ULX}

This source was first catalogued by Walton et al. (2011) as a potential ULX, with followup revealing that it lies in the eULX range. NGC 5907 ULX was reported and discussed further in Sutton et al. (2012; 2013), with X-ray spectral analysis revealing the presence of a possible break in the energy spectrum above $\sim$3 keV, a feature indicative of the ultraluminous state, while timing studies showed large-scale variability of up to $\sim$ an order of magnitude. This combination is more indicative of the sULXs than the HLXs; as a result, it has been suggested that this source may be a combination of the extreme end of the MsBH regime with possible super-Eddington accretion. In order to confirm this, further X-ray analysis is required along with multi-wavelength follow-up to confirm its distance, luminosity, spectral evolution \& companion type.

\section{Summary}

The study of ULXs began in the late '70s, and since that time, much has been learned about these fascinating objects, and yet, as with many other fields of research, many more questions still stand unanswered. In this paper, we have tried to briefly review our current knowledge, while exploring the new sub-classes that have emerged. 

It is thought that the majority of the \emph{standard} or \emph{low luminosity} ULXs are either sMBHs or MsBHs that are accreting at  or above the Eddington limit, but this theory has yet to be confirmed by direct mass measurements. Optical studies with the current and next generation of telescopes will allow us to test this theory, while providing greater constraints on their companions. Advancements in X-ray telescopes have  opened up other avenues of study, presenting an opportunity to test ideas on their accretion geometry. 

Hyperluminous X-ray sources are thought to be amongst the strongest candidates for IMBHs, and yet only two of these objects have been confirmed to date.  

\emph{Extreme} ULXs are the newest sub-class of these sources, and the group for which the least is known, yet it seems as though they may be a combination of the lower luminosity IMBHs and the most extreme MsBHs in our universe. 

For both eULXs and HLXs, more candidates must be identified and confirmed to achieve a greater understanding of these sub-classes. X-ray surveys are required to identify additional candidates and perform statistical analysis on such a sample, while multi-wavelength analyses can be used to explore their nature in much the same way as for the sULXs. 

As we look towards the next 50 years of X-ray astronomy, we find that there are still a large number of questions surrounding each of these new sub-classes, and much still to discover.

\begin{acknowledgements}

JCG would like to thank the organisers of the conference for their kind invitation to present and submit this review, she would also gratefully acknowledge funding from the Avadh Bhatia Fellowship and from an Alberta Ingenuity New Faculty Award. JCG would also like to thank some of her many colleagues and collaborators for their discussions, Ideas, suggestions and tireless efforts in the search to increase our understanding of these sources. This includes (but is not limited to); Timothy Roberts, Sean Farrell, Roberto Soria, Matthew Middleton, Fabien Gris{\'e}, Mar Mezcua, Hua Feng, Chris Done, Andrew Sutton, Matthew Servillat, Doug Swartz, Peter Jonker, Andreas Zezas, Manfred Pakull, Phil Kaaret, Tod Strohmayer, Lucy Heil, David Russell, Yi-Jung Yang, Andrew Goulding, Chris Copperwheat, Tom Maccarone, Andrew Levan, Mike Goad, and many others. 

\end{acknowledgements}

\bibliographystyle{aa}

\begin{thebibliography}{}

\bibitem{2002ApJ...568L..97B} Begelman M.~C., 2002, ApJ, 568, L97 

\bibitem[Bellovary et al.(2010)]{2010ApJ...721L.148B} Bellovary, J.~M., et al.\ 2010, \apjl, 721, L148 

\bibitem[Belczynski et al.(2010)]{2010ApJ...715L.138B} Belczynski, K., Dominik, M., Bulik, T., et al.\ 2010, \apjl, 715, L138 

\bibitem[Berghea et al.(2008)]{2008ApJ...687..471B} Berghea, et al., \ 2008, \apj, 687, 471 

\bibitem[1]{1999ApJ...519...89C} Colbert E.~J.~M., Mushotzky R.~F., 1999, ApJ, 519, 89 

\bibitem[Collura et al.(1994)]{1994ApJ...420L..63C} Collura, A., Reale, F., Schulman, E., \& Bregman, J.~N.\ 1994, \apjl, 420, L63 

\bibitem[Copperwheat et al. (2005)]{2005MNRAS.362...79C} Copperwheat C., et el., 2005, MNRAS, 362, 79 

\bibitem[Copperwheat et al. (2007)]{2007MNRAS.376.1407C} Copperwheat C., et al., 2007, MNRAS, 376, 1407 

\bibitem[Dewangan et al.(2006)]{2006ApJ...637L..21D} Dewangan, G.~C., Titarchuk, L., \& Griffiths, R.~E.\ 2006, \apjl, 637, L21 

\bibitem[Dheeraj \& Strohmayer(2012)]{2012ApJ...753..139D} Dheeraj, P.~R., \& Strohmayer, T.~E.\ 2012, \apj, 753, 139 

\bibitem[3]{2001ApJ...562L..19E} Ebisuzaki, T., et al.\ 2001, \apjl, 562, L19 


\bibitem[Farrell et al.(2011)]{2011AN....332..392F} Farrell, S.~A., et al.\ 2011, AN, 332, 392

\bibitem[Farrell et al.(2012)]{2012ApJ...747L..13F} Farrell, S.~A., et al.\ 2012, \apjl, 747, L13 

\bibitem[Feng et al.(2010)]{2010ApJ...710L.137F} Feng, H., Rao, F., \& Kaaret, P.\ 2010, \apjl, 710, L137 

\bibitem[Feng \& Kaaret(2010)]{2010ApJ...712L.169F} Feng, H., \& Kaaret, P.\ 2010, \apjl, 712, L169 

\bibitem[Feng \& Soria(2011)]{2011NewAR..55..166F} Feng, H., \& Soria, R.\ 2011, Nature, 55, 166 

\bibitem[Fiorito \& Titarchuk(2004)]{2004ApJ...614L.113F} Fiorito, R., \& Titarchuk, L.\ 2004, \apjl, 614, L113 

\bibitem[Forman et al.(1978)]{1978ApJS...38..357F} Forman, W., et al.\ 1978, \apjs, 38, 357 

\bibitem[Fryer \& Kalogera(2001)]{2001ApJ...554..548F} Fryer, C.~L., \& Kalogera, V.\ 2001, \apj, 554, 548 

\bibitem[Gladstone \& Roberts(2009)]{2009MNRAS.397..124G} Gladstone, J.~C., \& Roberts, T.~P.\ 2009, \mnras, 397, 124 

\bibitem[Gladstone et al.(2009)]{2009MNRAS.397.1836G} Gladstone, J.~C., Roberts, T.~P., \& Done, C.\ 2009, \mnras, 397, 1836 

\bibitem[Gladstone et al.(2013)]{2013ApJS..206...14G} Gladstone, J.~C., Copperwheat, C., Heinke, C.~O., et al.\ 2013, \apjs, 206, 14 

\bibitem{2003MNRAS.339..793G} Grimm H.-J., Gilfanov M., Sunyaev R., 2003, MNRAS, 339, 793 

\bibitem[Gris{\'e} et al.(2012)]{2012ApJ...745..123G} Gris{\'e}, F., Kaaret, P., Corbel, S., et al.\ 2012, \apj, 745, 123 

\bibitem[5]{2009MNRAS.397.1061H} Heil L.~M., Vaughan S., \& Roberts T.~P., 2009, MNRAS, 397, 1061 

\bibitem[Islam et al.(2004)]{2004MNRAS.354..427I} Islam, R.~R., Taylor, J.~E., \& Silk, J.\ 2004, \mnras, 354, 427 

\bibitem[Jonker et al.(2010)]{2010MNRAS.407..645J} Jonker, P.~G., et al.\ 2010, \mnras, 407, 645 

\bibitem[Jonker et al.(2012)]{2012ApJ...758...28J} Jonker, P.~G., et al.\ 2012, \apj, 758, 28 

\bibitem[8]{2005ChJAS...5..139K} Kaaret P., 2005, ChJAS, 5, 139 

\bibitem[Kaaret et al.(2006)]{2006ApJ...646..174K} Kaaret, P., Simet, M.~G., \& Lang, C.~C.\ 2006, \apj, 646, 174 

\bibitem[Kaaret \& Feng(2007)]{2007ApJ...669..106K} Kaaret, P., \& Feng, H.\ 2007, \apj, 669, 106 

\bibitem[Kaaret \& Corbel(2009)]{2009ApJ...697..950K} Kaaret, P., \& Corbel, S.\ 2009, \apj, 697, 950 

\bibitem[King et al.(2001)]{2001ApJ...552L.109K} King, A.~R., et al.,\ 2001, \apjl, 552, L109 

\bibitem[4]{2002A&A...382L..13K} K{\"o}rding E., Falcke H., Markoff S., 2002, A\&A, 382, L13

\bibitem[Lasota et al.(2011)]{2011ApJ...735...89L} Lasota, J.-P., et al.\ 2011, \apj, 735, 89 

\bibitem[2]{2001ApJ...551L..27M} Madau, P., Rees, M.~J.\ 2001, \apjl, 551, L27 

\bibitem[19]{2008ApJ...688.1235M} Madhusudhan N., et al., 2008, ApJ, 688, 1235 

\bibitem[Matsumoto et al.(2004)]{2004PThPS.155..379M} Matsumoto, H., Tatsuya, I., Tsuru, T.~G., et al.\ 2004, PThPS, 155, 379 

\bibitem[Middleton et al.(2011)]{2011MNRAS.411..644M} Middleton, M.~J., Roberts, T.~P., Done, C., \& Jackson, F.~E.\ 2011, \mnras, 411, 644 

\bibitem[Middleton et al.(2011a)]{2011MNRAS.417..464M} Middleton, M.~J., Sutton, A.~D., \& Roberts, T.~P.\ 2011a, \mnras, 417, 464 

\bibitem[56]{2003ApJ...585L..37M} Miller, J.~M., et al., 2003, ApJ, 585, L37 


\bibitem[Moon et al.(2011)]{2011ApJ...731L..32M} Moon, D.-S., et al.,\ 2011, \apjl, 731, L32

\bibitem[Okajima et al.(2006)]{2006ApJ...652L.105O} Okajima, T., Ebisawa, K., \& Kawaguchi, T.\ 2006, \apjl, 652, L105 

\bibitem[10]{2002astro.ph..2488P} Pakull M.~W., Mirioni L., 2002, astro, arXiv:astro-ph/0202488 

\bibitem[Piccinotti et al.(1982)]{1982ApJ...253..485P} Piccinotti, G., Mushotzky, R.~F., Boldt, E.~A., et al.\ 1982, \apj, 253, 485 

\bibitem[Pizzolato et al.(2010)]{2010MNRAS.406.1116P} Pizzolato, F., Wolter, A., \& Trinchieri, G.\ 2010, \mnras, 406, 1116 

\bibitem[Portegies Zwart et al.(2004)]{2004MNRAS.355..413P} Portegies Zwart, S.~F., Dewi, J., \& Maccarone, T.\ 2004, \mnras, 355, 413 


\bibitem[Roberts et al.(2002)]{2002MNRAS.337..677R} Roberts, T.~P., et al.,\ 2002, \mnras, 337, 677 

\bibitem{2006MNRAS.371.1877R} Roberts T.~P., et al., 2006, MNRAS, 371, 1877 

\bibitem[29]{2008MNRAS.387...73R} Roberts T.~P., Levan A.~J., Goad M.~R., 2008, MNRAS, 387, 73 

\bibitem[Roberts et al.(2011)]{2011AN....332..398R} Roberts, T.~P., et al.\ 2011, AN, 332, 398 

\bibitem[Schneider et al.(2002)]{2002ApJ...571...30S} Schneider, R., Ferrara, A., Natarajan, P., \& Omukai, K.\ 2002, \apj, 571, 30 

\bibitem[Servillat et al.(2011)]{2011ApJ...743....6S} Servillat, M., et al.\ 2011, \apj, 743, 6 

\bibitem[Soria et al.(2010)]{2010MNRAS.405..870S} Soria, R., et al.\ 2010, \mnras, 405, 870 

\bibitem[Soria et al.(2012)]{2012MNRAS.420.3599S} Soria, R., et al.,\ 2012, \mnras, 420, 3599 

\bibitem[5]{2006MNRAS.368..397S} Stobbart A.~M., Roberts T.~P., Wilms J., 2006, MNRAS, 368, 397 

\bibitem[Strohmayer \& Mushotzky(2003)]{2003ApJ...586L..61S} Strohmayer, T.~E., \& Mushotzky, R.~F.\ 2003, \apjl, 586, L61 

\bibitem[57]{2007ApJ...660..580S} Strohmayer T.~E., et al, 2007, ApJ, 660, 580

\bibitem[Sutton et al.(2012)]{2012MNRAS.423.1154S} Sutton, A.~D., et al.,\ 2012, \mnras, 423, 1154 

\bibitem[Sutton et al.(2013)]{2013} Sutton, A.~D., et al.,\ 2012, \mnras, \emph{\small{in press}} 

\bibitem[9]{2008ApJ...684..282S} Swartz D.~A., Soria R., Tennant A.~F., 2008, ApJ, 684, 282 

\bibitem[Swartz et al.(2011)]{2011ApJ...741...49S} Swartz, D.~A., Soria, R., Tennant, A.~F., \& Yukita, M.\ 2011, \apj, 741, 49 

\bibitem[Tsuru et al.(2004)]{2004PThPS.155...59T} Tsuru, T.~G., et al.\ 2004, PThPS, 155, 59 

\bibitem[Volonteri(2010)]{2010A&ARv..18..279V} Volonteri, M.\ 2010, \aapr, 18, 279 

\bibitem[Voss et al.(2011)]{2011MNRAS.418L.124V} Voss, R., et al.,\ 2011, \mnras, 418, L124 

\bibitem[Walton et al.(2011)]{2011MNRAS.416.1844W} Walton, D.~J., Roberts, T.~P., Mateos, S., \& Heard, V.\ 2011, \mnras, 416, 1844 


\bibitem[Yu \& Tremaine(2002)]{2002MNRAS.335..965Y} Yu, Q., \& Tremaine, S.\ 2002, \mnras, 335, 965 

\bibitem[Yuan et al.(2007)]{2007ApJ...658..282Y} Yuan, F., Taam, R.~E., Misra, R., Wu, X.-B., \& Xue, Y.\ 2007, \apj, 658, 282 

\bibitem{2009MNRAS.400..677Z} Zampieri L., Roberts T.~P., 2009, MNRAS, 400, 677 


\end{thebibliography}

\end{document}